\documentclass[12pt,a4paper]{article}\usepackage{a4wide}

\usepackage{epsfig}
\usepackage{amssymb,latexsym,amsmath}
\usepackage{amsfonts}
\usepackage{amsbsy}

\numberwithin{equation}{section}

\DeclareMathOperator{\tr}{tr}
\DeclareMathOperator{\rk}{rank}

\newcommand{\ui}{\textrm{i}}
\newcommand{\ue}{\textrm{e}}
\newcommand{\UI}{\textrm{I}}

\newcommand{\rz}{{\mathbb R}}

\newcommand{\kz}{{\mathbb C}}

\newcommand{\A}{{\mathbb A}}
\newcommand{\B}{{\mathbb B}}

\newcommand{\SM}{{\mathbb S}}
\newcommand{\T}{{\mathbb T}}

\newcommand{\CSE}{\textrm{CSE}}
\newcommand{\COE}{\textrm{COE}}
\newcommand{\SU}{\textrm{SU}}

\newtheorem{property}{Lemma}
\newtheorem{corollary}{Corollary}
\newtheorem{theorem}{Theorem}

\newcommand{\bdm}{\begin{displaymath}}
\newcommand{\edm}{\end{displaymath}}
\newcommand{\beq}{\begin{equation}}
\newcommand{\eeq}{\end{equation}}
\newcommand{\beqa}{\begin{eqnarray}}
\newcommand{\eeqa}{\end{eqnarray}}

\newcommand{\sep}{\setlength\arraycolsep{2pt}}

\newcommand{\prf}{\noindent{\bf Proof.}\phantom{X}}
\newcommand{\eprf}{\hfill$\Box$}

\begin{document}

\thispagestyle{empty}

\noindent ULM-TP/05-6\\November 2005
\vspace{2cm}

\begin{center}

{\LARGE\bf  The spectral form factor for quantum graphs}\\
\vspace*{5mm}
{\LARGE\bf with spin-orbit coupling}\\
\vspace*{2cm}
{\large Jens Bolte}%
\footnote{E-mail address: {\tt jens.bolte@uni-ulm.de}}

\vspace*{5mm}

Abteilung Theoretische Physik\\
Universit\"at Ulm, Albert-Einstein-Allee 11\\
D-89069 Ulm, Germany 

\vspace*{1cm}

{\large Jonathan Harrison}%
\footnote{E-mail address: {\tt jon@math.tamu.edu}}
\vspace*{5mm}

Department of Mathematics\\
Texas A\&M University\\
College Station\\
TX 77843-3368\\
USA

\end{center}

\vfill

\begin{abstract} 
We consider quantum graphs with spin-orbit couplings at the vertices.
Time-reversal invariance implies that the bond $S$-matrix is in the 
orthogonal or symplectic symmetry class, depending on spin quantum number
$s$ being integer or half-integer, respectively. The periodic-orbit 
expansion of the spectral form factor is shown to acquire additional weights 
from spin rotations along orbits. We determine the spin contribution to 
the coefficients in an expansion of the form factor from properties of 
the representation of the group of spin transformations on the graph. 
Consistency with the Circular Orthogonal and Circular Symplectic
Ensemble, respectively, of random matrices is obtained.
\end{abstract}


\newpage

\section{Introduction}
\label{intro}
Since their introduction in the field of quantum chaos by Kottos and
Smilansky \cite{paper:kottossmilansky2,paper:kottossmilansky}, quantum 
graphs have played an important role in efforts towards a deeper
understanding of correlations in spectra of classically chaotic quantum 
systems. According to the conjecture of Bohigas, Giannoni, and Schmit
\cite{paper:bohigasgiannonischmit}, these correlations can be described
by random matrix theory (RMT) \cite{book:mehta}. Although overwhelming 
evidence supports this conjecture, until recently the theoretical 
understanding of the RMT connection remained rather poor. 

Spectral two-point correlations are often measured in terms of the form 
factor $K(\tau)$. Its small-$\tau$ asymptotics have been the subject of 
recent 
studies, producing a considerably improved understanding of spectral 
correlations  in terms of correlations 
among classical periodic orbits
\cite{paper:sieberrichter,paper:sieber,paper:mulleretal}. 
In this context quantum graphs have proven 
ideal models for detailed investigations into the duality between
eigenvalue and periodic orbit correlations
\cite{paper:berkolaikoschanzwhitney,paper:berkolaikoschanzwhitney2,
paper:berkolaiko}. 

A guiding principle of the RMT conjecture is that symmetries of the
quantum system determine the universality class of spectral correlations. 
Usually, the presence or absence of time-reversal invariance in quantum 
systems with integer or half-integer total spin is taken as an indicator
whether an orthogonal, a unitary, or a symplectic universality class
is appropriate. Since Berry's pioneering 
semiclassical analysis of the form factor \cite{paper:berry}, 
however, a vast majority of 
investigations have concentrated on the orthogonal and the unitary case. 
(See \cite{paper:scharfdietzkushaakeberry,paper:boltekeppeler3,
paper:keppelermarklofmezzadri} for exceptions where the symplectic case 
is considered.) Quantum graphs with spin 1/2 were
introduced in \cite{paper:bolteharrison} in the context of a realization 
of a Dirac operator on graphs. We demonstrated that in this way a
quantum system in the symplectic universality class can be realized,
and analyzed the form factor in the light of the recent developments
\cite{paper:bolteharrison2}. 

Here we extend our previous studies of spin-orbit coupling on quantum
graphs to the case of arbitrary values of the spin quantum number $s$.
Since the use of Dirac operators is restricted to the case of $s=1/2$,
this extension is most conveniently performed with a Pauli operator. 
Starting from the case of spin zero, we introduce a spin-orbit coupling 
that is localized at the vertices, in terms of boundary conditions
describing spin rotations at the vertices. We then continue to 
investigate the form factor for the spectrum of the bond $S$-matrix 
for a Pauli operator, following the diagrammatic method introduced in
\cite{paper:berkolaikoschanzwhitney,paper:berkolaikoschanzwhitney2}.
We analyze the group of spin rotations on the graph generated by those
at the vertices and show that properties of its irreducible 
representations fix the spin contribution to the form factor.
In particular, it is found that a change from integer to half-integer
spin induces the same transformation of the form factor as exists between
the Circular Orthogonal Ensemble (COE) and Circular Symplectic Ensemble 
(CSE) in RMT.

This paper is organized as follows: After an introduction we
recall basic facts about graphs in section~\ref{1sec}. The construction
of Pauli operators on graphs and the structure of the corresponding
bond $S$-matrices is explained in section~\ref{2sec}. The definition
of the form factor and the removal of Kramers' degeneracy is presented
in section~\ref{s:form factor}. Section~\ref{s:spin contribution} is
then devoted to discussing the spin contribution to the form factor and 
in section~\ref{s:correlations} we evaluate the spin contribution.
Group theoretic properties of representations used in the calculation 
of the spin contribution are explained in an appendix.

\section{Graphs}
\label{1sec}
A compact graph $G$ consists of $V$ vertices connected by $B$ bonds.
The topology of $G$ is encoded in the connectivity matrix $C$. This
is a $V\times V$ matrix with entries
\begin{equation}
\label{conmatrix}
  C_{ij} := 
  \left\{ \begin{array}{cl}
          1 & \textrm{if vertices $i$ and $j$ are connected},\\
          0 & \textrm{otherwise}.\\
          \end{array} \right. 
\end{equation}
The valency $v_i$ of a vertex $i$ is the number of bonds meeting in the
vertex. Throughout we assume that $G$ possesses no loops so that the 
diagonal elements of $C$ vanish, and that any pair of vertices is connected 
by at most one bond. Hence $v_i = \sum_j C_{ij}$ and $2B = \sum_i v_i$.
Paths on $G$ are sequences $(b_1,\dots,b_t)$ of consecutive bonds,
and periodic orbits are periodic sequences in which case $t$ is the
period. We require the graph to be connected, i.e., any pair of vertices 
can be joined by a path. 

By assigning lengths $L_b$ to bonds $b$ we turn $G$ into a metric 
graph. On each bond $b$ we then introduce a coordinate $x_b \in [0,L_b]$.
The graph therefore becomes directed, since the coordinate on a bond
$b=(ij)$ connecting the vertices $i$ and $j$ runs from $i$ to $j$. 
In 
order to avoid degeneracies in the length spectrum of periodic orbits we
also demand that the lengths $L_b$ be rationally independent.
 
Functions $\psi=(\psi_1,\dots,\psi_B)$ on the graph can now be defined
in terms of functions $\psi_b : [0,L_b]\to\kz$ on the bonds. The quantisation
on graphs finally requires to introduce the Hilbert space
\begin{equation}
\label{L2space}
 L^2 (G) = \bigoplus_{b=1}^B L^2 (0,L_b) \ . 
\end{equation}
Further spaces of functions on $G$ are defined analogously. E.g.,
the components of $\psi\in W^{2,2}(G)$ are functions in the 
$L^2$-Sobolev spaces $W^{2,2}(0,L_b)$. 
\section{Pauli operators on graphs}
\label{2sec}
Usually metric, compact graphs $G$ are quantized in terms of a 
suitable realization of the Laplacian on $L^2(G)$. One thus 
describes a point-like quantum particle moving freely along the bonds 
of the graph, with local interactions at the vertices. These interactions 
are described in terms of the boundary conditions that specify a given 
self-adjoint realization of the Laplacian. We will closely follow the 
method of Kostrykin and Schrader \cite{paper:kostrykinschrader}, which we 
briefly now recall: As a differential expression the Laplacian reads 
$\Delta\psi = (\psi_1'',\dots,\psi_B'')$. Any self-adjoint realization
of this operator can be specified in terms of two complex 
$2B\times 2B$ matrices $\A$ and $\B$, when $\rk (\A,\B)=2B$ and $\A\B^\dagger$ 
is hermitian. In such a case an operator core consists of  
$\psi\in W^{2,2}(G)$ fulfilling
\begin{equation}
\label{bc}
 \A \boldsymbol{\psi} + \B \boldsymbol{\psi}' = 0 \ .
\end{equation}
Here $\boldsymbol{\psi}$ denotes the vector of the $2B$ boundary values 
of $\psi$, whereas $\boldsymbol{\psi}'$ is the corresponding vector of 
inward derivatives at the vertices. In order for the boundary conditions 
to be local, we furthermore require a block structure of the matrices $\A$ 
and $\B$, such that only boundary values at the same vertex are related 
through (\ref{bc}).

The spectrum of $-\Delta$ on the graph is discrete, non-negative and has no 
finite accumulation point. Following Kottos and Smilansky
\cite{paper:kottossmilansky2,paper:kottossmilansky} this spectrum can most 
conveniently be characterized in terms of the so-called bond S-matrix 
$\SM(k)$, where $k\in\rz$ is such that $\lambda =k^2$ is the spectral 
parameter of $-\Delta$: $k^2$ is an eigenvalue, iff $k$ is 
a solution of
\begin{equation}
\label{detS}
 \det \bigl( \UI_{2B} - \SM(k) \bigr) = 0 \ .
\end{equation}
The bond S-matrix is a unitary, $2B\times 2B$ matrix defined in terms 
of the local transition matrices $\T^{(i)}(k)$ that reflect the boundary 
conditions at the vertex $i$ prescribed by the blocks $\A^{(i)}$ and
$\B^{(i)}$ of $\A$ and $\B$, respectively,
\begin{equation}
\label{vertexs-m}
 \T^{(i)}(k) = -\bigl( \A^{(i)} + \ui k \, \B^{(i)} \bigr)^{-1} 
               \bigl(\A^{(i)} - \ui k \, \B^{(i)}) \ .
\end{equation}
More precisely, the entries of $\SM(k)$, labeled by the bonds $(ij)$
connecting the vertices $i$ and $j$, read
\begin{equation}
\label{bonds-m}
 S^{(ij)(lm)} = 
 \delta_{im}\,T^{(i)}_{(ij)(lm)}\,\ue^{\ui k L_{(lm)}} \ .
\end{equation}
Time reversal is implemented by complex conjugation, so that a 
time-reversal invariant realization of the Laplacian is obtained when 
$\A$ and $\B$ are real. This implies that besides being unitary the bond 
$S$-matrix is symmetric, and hence possesses the symmetries of the Circular 
Orthogonal Ensemble (COE) of RMT.

We now introduce spin-orbit coupling on quantum graphs following the 
non-relativistic Pauli equation. In a first step we therefore consider
the Laplacian realized on $L^2(G)\otimes\kz^{2s+1}$, where 
$s\in\{1/2,1,3/2,\dots\}$ denotes the spin quantum number. As long as
the boundary conditions are taken over from the previous case without 
spin, through a trivial extension of the condition (\ref{bc}) to the $2s+1$ 
components of the functions on each bond, no spin-orbit coupling is present. 
A local spin-orbit interaction at the vertices can now be introduced in a 
very simple way: one merely has to allow for general $2B(2s+1)\times 2B(2s+1)$ 
matrices $\A$ and $\B$, with $\rk (\A,\B)$ maximal and $\A\B^\dagger$ 
hermitian, 
to determine boundary conditions in analogy to (\ref{bc}). Locality 
furthermore requires, as above, that these condition be satisfied at each 
vertex separately.

For a spin-$s$ system the time reversal operator $T_s$ is anti-unitary, 
with $T_s^2 = (-1)^{2s}$, see \cite{paper:wigner}. In the case of integer 
spin, requiring time-reversal invariance of the Pauli operator therefore 
amounts to the same condition as in the case of the Laplacian: the bond 
$S$-matrix has to be unitary and symmetric. For half-integer spin, however, 
time-reversal invariance implies Kramers' degeneracy \cite{paper:kramers}, 
i.e., a twofold degeneracy in the spectrum of the Pauli operator. Moreover, 
the bond $S$-matrix possesses the symmetries of the Circular Symplectic 
Ensemble (CSE), see \cite{book:mehta,book:haake}. The same conditions apply 
for the Dirac operator on a graph, corresponding to $s=1/2$. This case was 
studied in \cite{paper:bolteharrison}. In the general case of  
spin $s$ one merely has to replace the $\SU(2)$ matrices $u$ that provide 
the spin rotations at the vertices by their $2s+1$-dimensional unitary 
irreducible representations $R^s (u)$. Requiring invariance of the 
transition matrix (\ref{vertexs-m}) under a permutation of the bonds at 
the vertex as in \cite{paper:bolteharrison}, this yields
\begin{equation}
\label{eq:3gen}
 \T^{(i)} = \left( \begin{array}{ccc} R^s (u_{1}) & & \\
            & \ddots & \\
            & & R^s (u_{v_{i}}) \\
            \end{array} \right)
            \left( X \otimes \UI_{2s+1} \right)
            \left( \begin{array}{ccc}
            R^s (u_{1})^{-1} & & \\
            & \ddots & \\
            & & R^s (u_{v_{i}})^{-1} \\
\end{array} \right) \ .
\end{equation}
Here $X$ is a $v_i\times v_i$ matrix that is required to be unitary and
invariant under simultaneous permutations of columns and rows. These
conditions imply that this matrix must be of the form,
\begin{equation}
\label{eq:3X}
 X = \ue^{\ui \theta}
     \left( \begin{array}{ccc}
     q -1 & & q \\
     & \ddots & \\
     q & & q -1 \\
     \end{array} \right) 
 \quad \text{with} \quad q :=\frac{1+\ue^{\ui \omega}}{v_{i}} \ ,
\end{equation}
where $\omega$ and $\theta$ are real parameters. It is easy to see that
$\omega=\pi$ corresponds to Dirichlet boundary conditions on the bonds.
The most common choice, however, is $\omega = 0$, leading to Neumann 
boundary conditions \cite{paper:kottossmilansky2,paper:kottossmilansky}.
The bond $S$-matrix constructed from (\ref{eq:3gen}) and (\ref{eq:3X})
according to (\ref{bonds-m}) therefore yields, for fixed $k$, the general 
form of a unitary matrix that may serve as an $S$-matrix of a quantum
graph with spin-orbit coupling.
\section{The form factor}\label{s:form factor}
The purpose of this paper is to calculate the effect of spin in one 
particular, 
commonly studied, spectral statistic of a quantum graph, 
the form factor.  We follow
the approach taken in
\cite{paper:berkolaikoschanzwhitney, paper:berkolaikoschanzwhitney2}
and consider the form factor derived from the spectrum of the $S$-matrix.
For the $S$-matrix spectrum we replace $\ue^{\ui k L_{(i j )}}$ in 
(\ref{bonds-m}) with $\ue^{\ui \phi_{(i j )}}$. 
The $B$ phases $\phi_{( ij ) }$ are random variables uniformly 
distributed in $[ 0 , 2 \pi ]$.  They define an ensemble of matrices 
$\SM_{\phi}$ over which we average, equivalent to averaging over bond 
lengths.
With such a replacement the $S$-matrix is   
\begin{equation}\label{S-matrix}
S_{\phi}^{(ij)(kl)}:=\delta_{il}\,  \sigma_{(ij)(ki)} \, R^{s}\big( 
u^{(ij)(ki)} \big) 
\, \ue^{\ui \phi_{(kl)}} \ .
\end{equation}
The Kronecker-delta ensures transitions only occur 
between bonds connected at 
a vertex.
$\sigma_{(ij)(ki)}$ is the $(ij),(ki)$ element of the matrix $X$
in (\ref{eq:3gen}) and 
$R^{s} (u^{(ij)(ki)})$ is a spin-$s$ representation of an element of 
$\SU(2)$ describing the spin transformation at the vertex $i$.  
According to (\ref{eq:3gen}),
\begin{equation}\label{spin transform}
R^s\big(u^{(ij)(ki)}\big)= 
R^s\big( u^{(i)}_{j} \big) R^s\big( u^{(i)}_{k}\big)^{-1} \ .
\end{equation}
  
Having defined the $S$-matrix, the form factor is introduced as in 
\cite{book:haake}. In the case of integer spin the $S$-matrix generically 
has $N=2B(2s+1)$ non-degenerate eigenvalues. Thus
\begin{equation}\label{eq:form factor orth}
K_{\mathrm{orth}}(\tau_{\mathrm{orth}}) := 
\frac{1}{N} \left\langle | \tr \SM_{\phi}^{t} |^{2} 
\right\rangle_{\phi} \ , \qquad \tau_{\mathrm{orth}}=\frac{t}{2B(2s+1)} \ .
\end{equation}
For half-integer spin the form factor is defined after first removing Kramers'
degeneracy as explained in \cite{paper:boltekeppeler3}. The $S$-matrix then 
has $N=B(2s+1)$ independent eigenvalues, and 
\begin{equation}\label{eq:form factor sympl}
K_{\mathrm{sympl}} (\tau_{\mathrm{sympl}})
:= \frac{1}{4N} \left\langle | \tr \SM_{\phi}^{t} |^{2} 
\right\rangle_{\phi} \ , \qquad 
\tau_{\mathrm{sympl}}=\frac{t}{B(2s+1)} \ .
\end{equation}
We distinguish the different definitions of 
the form factors and parameters $\tau$
with labels corresponding to the symmetry introduced 
by time-reversal invariance.

Expanding the trace of $\SM^{t}$ as a sum over the set $P_t$ of 
periodic orbits of period $t$ yields
\begin{equation}\label{trace expansion}
\tr \SM_{\phi}^t = \sum_{p\in P_t} \frac{t}{r_{p}} \, A_{p} \, 
\ue^{\ui\pi\mu_p} \, \tr \big( R^s (d_{p})\big) \, \ue^{\ui \phi_{p}} \ .
\end{equation}
The periodic orbit $p$ consists of a series of subsequently visited bonds 
$(b_1 , b_2 , \dots , b_t )$. It has associated with it the quantities
\begin{equation}
\begin{split}
A_p \, \ue^{\ui\pi\mu_p} 
       &:=\sigma_{b_t b_{t-1}} \sigma_{b_{t-1} b_{t-2} }
\dots \sigma_{b_2 b_1} \ , \\
d_p    &:=u^{b_t b_{t-1}} u^{b_{t-1} b_{t-2}} \dots u^{b_2 b_1} \ , \\
\phi_p & :=\sum_{j=1}^{t} \phi_{b_j} \ .
\end{split}
\end{equation}
The phases $\mu_p$ are such that $A_p >0$, and $r_{p}$ is the repetition 
number of $p$. $t/r_p$ is the number of equivalent starting positions 
of an orbit due to cyclic permutations.

Using the periodic orbit expansion,
\begin{equation}\label{sym form factor}
\left\langle | \tr \SM_{\phi}^{t} |^{2} 
\right\rangle_{\phi} = t^2 \sum_{p,q \in P_{n}} 
\frac{A_{p} A_{q}}{r_p r_q} \, \ue^{\ui\pi(\mu_p-\mu_q)} \, 
\chi_{R^s}(d_p) \chi_{R^s}^{*}(d_q) \, \delta_{\phi_p ,\phi_q } \ ,  
\end{equation}
where we have introduced character notation for the trace of a representation,
$\chi_{R^s}(d)=\tr R^s(d)$.
The Kronecker-delta results from averaging over the phases $\phi$.
It fixes contributing terms in the double sum to pairs of orbits in which 
each bond is visited the 
same number of times.  On a metric graph with rationally independent bond 
lengths this is equivalent to requiring the lengths of $p$ and $q$ be equal.

For comparison the form factor of a graph quantized with the Laplacian 
(spin-$0$), as studied in 
\cite{paper:berkolaikoschanzwhitney, paper:berkolaikoschanzwhitney2}, is
\begin{equation}\label{zero form factor}
K_{\mathrm{zero}}(\tau_{\mathrm{zero}}) := \frac{1}{2B}
\left\langle | \tr \SM_{\phi}^{t} |^{2} 
\right\rangle_{\phi} = \frac{t^2}{2B} 
\sum_{p,q \in P_{n}} \frac{A_{p} A_{q}}{r_p r_q} \, \ue^{\ui\pi(\mu_p-\mu_q)}  
\, \delta_{\phi_p ,\phi_q } \ ,
\end{equation}
with $\tau_{\mathrm{zero}}=t/2B$.

It was pointed out in \cite{paper:heusler} that 
the power series expansions of the CSE and COE form factors are 
closely connected, if one performs the same substitution that leads from
(\ref{eq:form factor orth}) to (\ref{eq:form factor sympl}),
\begin{equation}
\begin{split}
K_{\CSE}(\tau) & =  \frac{\tau}{2} + \frac{\tau^2}{4} + \frac{\tau^3}{8}
+ \frac{\tau^4}{12} + \dots \ ,\\
\frac{1}{2} K_{\COE}\left( \frac{\tau}{2} \right) 
& =  \frac{\tau}{2} - \frac{\tau^2}{4} + \frac{\tau^3}{8}
- \frac{\tau^4}{12} + \dots \ .
\end{split}
\end{equation}
Calling $K^{m}$ the term containing $\tau^m$ the relationship may be 
written
\begin{equation}\label{eq:RMT relation}
K^{m}_{\CSE}(\tau) = \left( -\frac{1}{2} \right)^{m+1} K^{m}_{\COE}(\tau) \ .
\end{equation}
According to the conjecture of Bohigas, Giannoni, and Schmit 
\cite{paper:bohigasgiannonischmit} in the semiclassical limit we expect the 
form factors of quantum graphs to correspond to those of random matrices.  
In particular,
\begin{equation}\label{form factor relation}
K^{m}_{\mathrm{sympl}}(\tau) = \left( -\frac{1}{2} \right)^{m+1} 
K^{m}_{\mathrm{orth}}(\tau) \ .
\end{equation}
It is this relation we propose to establish in quantum graphs.

\section{Spin contributions to the form factor}\label{s:spin contribution}
The form factor is usually studied in the semiclassical 
limit which for quantum graphs corresponds to $B\rightarrow \infty$.  
For small but finite $\tau=t/N$ the limit of long orbits, 
$t\rightarrow \infty$, is also required. For details see 
\cite{paper:berkolaikoschanzwhitney, paper:berkolaikoschanzwhitney2}.
In this limit the proportion of orbits $p$ with $r_p \ne 1$ 
tends to zero and these orbits can effectively be ignored in equations 
(\ref{sym form factor}) and (\ref{zero form factor}).  Following
\cite{paper:berkolaikoschanzwhitney2} the sum over orbit pairs is organized 
in terms of diagrams.  A diagram consists of all pairs of orbits 
related by the same pattern of 
permutations of arcs between self-intersections
and time-reversal of arcs.   Consequently such pairs of orbits 
have identical phases $\phi_p=\phi_q$.  
Figures \ref{fig:singleint} and 
\ref{fig:doubleint} provide examples of diagrams.

We define the contribution to the form factor from a specific diagram $D$ 
with $n$ self-intersections to be $K^{n,D}$,{\sep
\beqa\label{eq:Km orth}
K^{n,D}_{\mathrm{orth}} (\tau_{\mathrm{orth}}) 
&:= &\frac{t^2}{2B(2s+1)} \sum_{(p,q) \in D_t}
A_{p} A_{q} \, \ue^{\ui\pi(\mu_p-\mu_q)} \, \chi_R(d_p) \chi_R^*(d_q) \ ,\\
\label{eq:Km sympl}
K^{n,D}_{\mathrm{sympl}} (\tau_{\mathrm{sympl}} ) 
&:= &\frac{t^2}{4B(2s+1)} \sum_{(p,q) \in D_t}
A_{p} A_{q} \, \ue^{\ui\pi(\mu_p-\mu_q)} \, \chi_R(d_p) \chi_R^*(d_q) \ . 
\eeqa
}Here $D_{t}$ is the set of pairs of orbits $(p,q)$ of period 
$t$ contained in $D$.

To separate spin contributions from the sum we assume that
the elements $R(d_p)$ are chosen randomly 
(independent of $p$) from a representation $R(\Gamma)$ of a subgroup 
$\Gamma \subseteq \SU(2)$.  
This can be achieved by selecting the 
matrices $R(u^{(i)}_j)$ randomly from $R(\Gamma)$.  Then
\begin{equation}\label{eq:sym 2}
\begin{split}
K^{n,D}_{\mathrm{orth}/\mathrm{sympl}}(\tau_{\mathrm{orth}/\mathrm{sympl}}) = &
\frac{\alpha_{\mathrm{orth}/\mathrm{sympl}}}{(2s+1)}
\left( \frac{1}{| D_{t}|} \sum_{(p,q) \in D_{t}} \chi_R(d_p) \chi_R^*(d_q) 
\right)\\ &\times \left( \frac{t^2}{2B}
\sum_{(p,q) \in D_{t}} A_{p} A_{q} \, \ue^{\ui\pi(\mu_p-\mu_q)} \right) \ ,
\end{split}
\end{equation}
where $\alpha_{\mathrm{orth}}=1$ and $\alpha_{\mathrm{sympl}}=1/2$.
The second term in (\ref{eq:sym 2}) is the contribution to the 
form factor of the graph with spin zero (\ref{zero form factor}),
\begin{equation}\label{eq:sym 3}
K^{n,D}_{\mathrm{zero}}  (\tau_{\mathrm{zero}}):=  \frac{t^2}{2B}
\sum_{(p,q) \in D_{t}} A_{p} A_{q} \, \ue^{\ui\pi(\mu_p-\mu_q)} \ .
\end{equation}

For a finite subgroup $\Gamma$ the matrices $R\big(u^{(i)}_j\big)$ 
are chosen independently with uniform probability 
$1/|\Gamma|$, where $|\Gamma|$ is the order of the subgroup. 
In the semiclassical limit the number and period of periodic orbits 
tends to infinity.  Consequently,
\beq
\frac{1}{| D_{t}|} \sum_{(p,q) \in D_{t}} \chi_R(d_p) \chi_R^*(d_q)
\to \frac{1}{|\Gamma{}|^t} 
\sum_{u_{b_1} \in \Gamma{}}\ldots \sum_{u_{b_t}\in \Gamma{}}
\chi_R(d_p) \chi_R^*(d_q) \ .
\eeq
If the subgroup $\Gamma \subseteq \SU(2)$ is continuous instead of
finite the sums over $\Gamma$ are replaced with integrals over the subgroup.
The elements $u^{(i)}_j\in \Gamma$ are then chosen randomly with 
respect to Haar measure on the subgroup.

We will show in section~\ref{s:correlations} that for $R$ an irreducible 
representation of $\Gamma$ of dimension $2s+1$,
\beq
\frac{1}{|\Gamma{}|^t} \sum_{u_{b_1}\in \Gamma{}} 
\ldots \sum_{u_{b_t}\in \Gamma{}}
\chi_R(d_p) \chi_R^*(d_q) =
\left( \frac{c_R}{2s+1} \right)^{n} \ , 
\eeq
where $n$ is the number of self-intersections at which arcs of the orbit 
$p$ have been rearranged to produce $q$.
The constant $c_R$ takes the value $c_R=1$ if the representation $R$ is real,
and $-1$ if it is quaternionic. 
Consequently contributions to the form factor have the form,
\beq
K^{n,D}_{\mathrm{orth}/\mathrm{sympl}}
(\tau{}_{\mathrm{orth}/\mathrm{sympl}}) = 
\frac{\alpha_{\mathrm{orth}/\mathrm{sympl}}}{(2s+1)}
\left( \frac{c_R}{2s+1} \right)^{n} 
K^{n,D}_{\mathrm{zero}}(\tau{}_{\mathrm{zero}})  \ . 
\eeq

Let us first consider integer $s$, in this case the dimension of $R$ is odd. 
Quaternionic representations can only occur with even dimension
(see \cite{book:hamermesh} section 5-5) and therefore 
$c_R=1$.  
For integer spin we use the form factor
(\ref{eq:form factor orth}). Hence
\beq
K^{n,D}_{\mathrm{orth}}\left( \frac{t}{2B(2s+1)} \right) 
= \frac{1}{(2s+1)^{n+1}} K^{n,D}_{\mathrm{zero}} 
\left( \frac{t}{2B} \right) \ .
\eeq
Conjecturing that $K^m(\tau)$, the term in the form factor expansion 
containing $\tau^m$, is generated by diagrams reordered at $n=m-1$ self 
intersections, we have
\beq
K^{m,D}_{\mathrm{orth}} ( \tau{}_{\mathrm{orth}}) 
= K^{m,D}_{\mathrm{zero}} ( \tau{}_{\mathrm{orth}}) \ .
\eeq
Introducing any irreducible representation of spin transformations 
for integer spin, the terms in the expansion of the form factor are
the same as those for the graph quantized with spin zero.

If we consider half-integer spin the spectrum of $\SM_\phi$ generically is 
doubly degenerate and the correct formula for the form factor is 
(\ref{eq:form factor sympl}), i.e.,
\beq
K^{n,D}_{\mathrm{sympl}} \left( \frac{t}{B(2s+1)} \right) 
= \frac{c_R^{n}}{2(2s+1)^{n+1}} 
K^{n,D}_{\mathrm{zero}} \left( \frac{t}{2B} \right) \ .
\eeq
Again, if we conjecture that the terms in the expansion 
containing $\tau^m$ are generated by diagrams with $n=m-1$ self-intersections
we find
\beq
K^{m,D}_{\mathrm{sympl}} (\tau_{\mathrm{sympl}}) = 
\frac{c_R^{m-1}}{2^{m+1}} K^{m,D}_{\mathrm{zero}} ( \tau_{\mathrm{sympl}}) \ .
\eeq
When $R$ is an irreducible quaternionic representation, $c_R=-1$,
this would establish the same relationship between the expansion 
of the form factor for a system with half-integer spin and the graph
quantized with spin zero as exists between the expansions 
of $K_{\CSE}(\tau)$ and $K_{\COE}(\tau)$, see (\ref{eq:RMT relation}).

To conclude, an example of a finite subgroup $\Gamma$ of spin transformations 
are Hamilton's quaternions, $\Gamma=\{ \pm \UI, \pm \ui \sigma_{x}, 
\pm \ui \sigma_{y},\pm \ui \sigma_{z} \}$,
%
%
where $\sigma_j$ is a Pauli matrix. $\Gamma$ is itself a two dimensional 
irreducible quaternionic representation corresponding to transformations 
of spin-$1/2$. In \cite{paper:keppelermarklofmezzadri} spin transformations 
from this subgroup are applied to the cat map and CSE statistics are 
observed.  As $\Gamma$ is both irreducible and quaternionic CSE statistics 
are expected even with spin transformations taken from such a small subgroup 
of $\SU(2)$.

\section{Spin correlations}\label{s:correlations}

%
%
Let $d_p= u_{b_{t}}u_{b_{t-1}} \dots u_{b_2} u_{b_1}$ be a product of 
elements of $u_{b_j}\in\Gamma$,
which corresponds to a periodic orbit $p=(b_1,b_2,\dots,b_t)$.
Here 
$u_{(ij)}=\big(u^{(i)}_{j}\big)^{-1}u^{(j)}_i=u_{(ji)}^{-1}$ is the element 
of $\SU(2)$ that transforms spin when traversing the bond $(ij)$. This 
labeling of elements of $\SU(2)$ with bonds simplifies calculations.
Let $d_q \in \Gamma$ be obtained from $d_p$ by taking the bonds 
of the periodic orbit $p$ in a different order.  This corresponds to 
permuting and inverting the elements $u_{b_j}$ appropriately.
\begin{theorem}\label{maintheorem}
\bdm
\frac{1}{|\Gamma{}|^t}
\sum_{u_{b_1} \in \Gamma{}}\dots \sum_{u_{b_t} \in \Gamma{}}
\chi_{R}(d_p) \chi_{R}^*(d_q) = \left( \frac{c_R}{2s+1} \right)^{n} \ ,
\edm
where $n$ is the number of self-intersections at which the orbit $p$ 
has been rearranged to produce $q$.
\end{theorem}
\prf
We will apply properties of group representations to calculate
average values of the product of the characters of $d_p$ and $d_q$.  The 
properties we use, together with proofs where appropriate, are given in 
the appendix.

To begin, if the orbit $q=p$ we have $n=0$. In this case
\beq
\begin{split}
 \frac{1}{|\Gamma{}|^t}\sum_{u_{b} \in \Gamma{}}
 \chi_{R}(d_p) \chi_{R}^*(d_p) 
 &= \frac{1}{|\Gamma{}|^t}\sum_{u_{b} \in \Gamma{}}
    \chi_{R}(u_{b_{t}}\dots u_{b_1}) \chi_{R}^*(u_{b_{t}}\dots 
    u_{b_1})\\
 &= \frac{1}{|\Gamma|}\sum_{u\in\Gamma}\chi_{R}(u)\chi_{R}^*(u)
   = 1 \ .
\end{split}
\eeq
The sum over $u_b \in \Gamma$ is an abbreviation for the sum over
all $u_{b_1},\dots,u_{b_t}$.  The sum was evaluated by a simple change of 
variables $u=u_{b_{t}}\dots u_{b_1}$ and character orthogonality.
%
%
%
%
The rest of the proof is inductive.  We consider two cases, firstly where 
$p$ and $q$ differ by reversing the direction of a section of $q$, and 
secondly where two sections of $q$ are permuted.\\

\noindent \textbf{1. Reordering at a single intersection.} 

\begin{figure}[htb]
\begin{center}
\includegraphics[width=10cm]{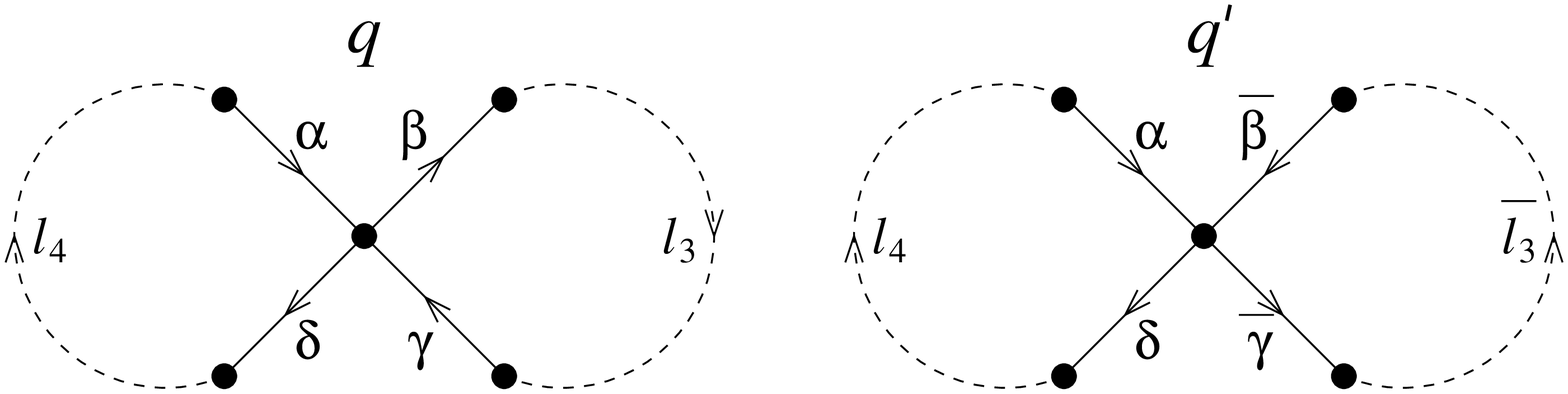}
\end{center}
\caption{The two orbits $q$ and $q'$ related by a change in the order of 
arcs at a single self-intersection.}
\label{fig:singleint}
\end{figure}

\noindent We assume the theorem holds for two orbits 
$p=(\alpha, \beta, l_1, \gamma, \delta, l_2)$ and 
$q=(\alpha, \beta, l_3, \gamma, \delta, l_4)$.
%
%
%
The notation follows Figure~\ref{fig:singleint}, showing orbit $q$;
$l_j$ denotes a loop of the orbit containing an unspecified number of bonds.

The order and direction in which bonds are traversed in the loops 
may differ between $p$ and $q$.  The number of self-intersections at 
which the orbit $q$ differs from $p$ is $n$ by assumption.  
We will show that the theorem also holds when applied 
to an orbit $q'=(\alpha, \overline{\gamma{}}, \overline{l_3}, 
\overline{\beta}, \delta, l_4)$,
%
%
where $\overline{b}$ describes the 
bond or loop being traversed in the opposite direction.  Figure 
\ref{fig:singleint} shows the relationship between $q$ and $q'$.
By definition $n'=n+1$, the order of bonds in $q'$ has been changed from 
the order in $q$ at a single self-intersection. 
Using $u_{\overline{b}}=u_b^{-1}$ and the definition of the orbit $q'$
we find
\beq
\begin{split}
 \frac{1}{|\Gamma{}|^t}\sum_{u_{b} \in \Gamma{}}
 \chi_{R}(d_p) \chi_{R}^*(d_{q'}) 
 &= \frac{1}{|\Gamma{}|^t}\sum_{u_{b} \in \Gamma{}}
    \chi_{R}(u_{l_2} u_{\delta} u_{\gamma} u_{l_1} u_{\beta} u_{\alpha} ) 
    \chi_{R}^*(u_{l_4} u_{\delta} u_{\beta}^{-1} u_{l_3}^{-1} 
    u_{\gamma{}}^{-1} u_{\alpha} ) \\
 &= \frac{1}{|\Gamma{}|^t}\sum_{u_{b} \in \Gamma{}}
    \chi_{R}(u_{l_2} u_y u_{l_1} u_x ) 
    \chi_{R}^*(u_{l_4} u_y u_z u_{l_3}^{-1} u_z u_x ) \ .
\end{split}
\eeq
Here $u_{l_j}$ is the product of elements $u_b$ picked up along the loop $l_j$,
and $u_x=u_\beta u_\alpha$, $u_y=u_\delta u_\gamma$, 
$u_z=u_\gamma^{-1} u_\beta^{-1}$. 
%
Corollary \ref{property1} (see the appendix) can now be used to evaluate 
the sum on $u_z$,
\beq
\frac{1}{|\Gamma{}|^t}\sum_{u_{b} \in \Gamma{}}
\chi_{R}(d_p) \chi_{R}^*(d_{q'}) = \left( \frac{c_R}{2s+1} \right)
\frac{1}{|\Gamma{}|^{t-1}}\sum_{u \in \Gamma{}}
\chi_{R}(u_{l_2} u_y u_{l_1} u_x )  
\chi_{R}^{*}(u_{l_4} u_y u_{l_3} u_x ) \ .
\eeq
Returning to the original variables we find,
\beq\label{eq:single}
\frac{1}{|\Gamma{}|^t}\sum_{u_{b} \in \Gamma{}}
\chi_{R}(d_p) \chi_{R}^*(d_{q'}) = \frac{c_R}{2s+1}
\frac{1}{|\Gamma{}|^{t}}\sum_{u \in \Gamma{}}
\chi_{R}(d_p) \chi_{R}^*(d_{q}) \ ,
\eeq
where we have introduced an extra dummy variable to account for 
the sum on $u_z$ which was equivalent either to summing on 
$u_\beta$ or $u_\gamma$.  Equation (\ref{eq:single}) shows that 
changing the orbit $q$ by reordering the bonds at a single self-intersection
introduces a factor $\frac{c_R}{2s+1}$. \\

\noindent \textbf{2. Reordering at a pair of self intersections.}

\begin{figure}[htb]
\begin{center}
\includegraphics[width=8cm]{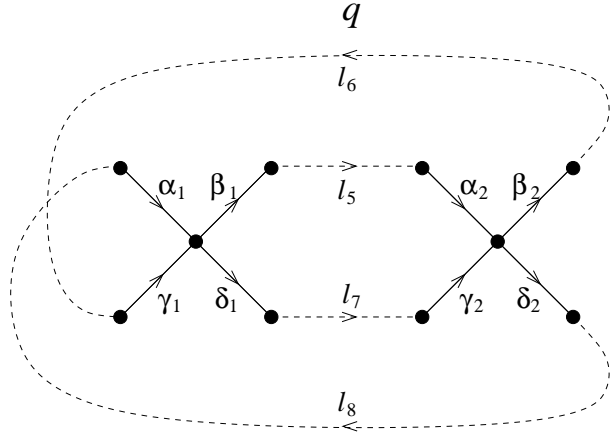}
\end{center}
\caption{An orbit $q$ with two self-intersections.}
\label{fig:doubleint}
\end{figure}

\noindent The second way to reorder bonds in the orbit $q$ is to permute two 
sections of the orbit. We consider a pair of orbits $p$ and $q$, where
\beq
\begin{split}
p&=
(\alpha_1, \beta_1, l_1, \alpha_2, \beta_2, l_2, 
\gamma_1, \delta_1, l_3,\gamma_2, \delta_2, l_4) \ , \\
q&=
(\alpha_1, \beta_1, l_5, \alpha_2, \beta_2, l_6, 
\gamma_1, \delta_1, l_7, \gamma_2, \delta_2, l_8) \ .
\end{split}
\eeq
Figure~\ref{fig:doubleint} shows the orbit $q$.

The orbit $q'$ is defined by exchanging two sections of $q$,
\beq
q'=
(\alpha_1, \delta_1, l_7,\gamma_2, \beta_2, l_6, 
\gamma_1, \beta_1, l_5, \alpha_2, \delta_2, l_8) \ .
\eeq
This changes the order of the arcs at two self-intersections. Thus, 
$n'=n+2$, and
\beq
\begin{split}
 \frac{1}{|\Gamma{}|^t}\sum_{u_{b} \in \Gamma{}}
 \chi_{R}(d_p) \chi_{R}^*(d_{q'}) 
 &= \frac{1}{|\Gamma{}|^t}\sum_{u_{b} \in \Gamma{}}\chi_{R}(u_{l_4}
     u_{\delta_2}u_{\gamma_2}u_{l_3} u_{\delta_1}u_{\gamma_1} u_{l_2}
     u_{\beta_2}u_{\alpha_2}u_{l_1}u_{\beta_1}u_{\alpha_1}) \\
 &\hspace{2cm}\times \chi_{R}^* (u_{l_8}u_{\delta_2}u_{\alpha_2}u_{l_5}
              u_{\beta_1}u_{\gamma_1}u_{l_6}u_{\beta_2}u_{\gamma_2}u_{l_7}
              u_{\delta_1}u_{\alpha_1}) \\
 &= \frac{1}{|\Gamma{}|^t}\sum_{u_{b} \in \Gamma{}}\chi_{R}(u_{l_4}u_{y_2}
    u_{l_3}u_{y_1}u_{l_2}u_{x_2}u_{l_1}u_{x_1}) \\
 &\hspace{2cm}\times\chi_{R}^* (u_{l_8}u_{y_2}u_{z_2}u_{l_5}u_{x_1}u_{z_1}^{-1}
              u_{l_6}u_{x_2}u_{z_2}^{-1}u_{l_7}u_{y_1}u_{z_1}) \ ,
\end{split}
\eeq
where $u_{x_j}=u_{\beta_j}u_{\alpha_j}$, $u_{y_j}=u_{\delta_j}u_{\gamma_j}$ 
and $u_{z_j}=u_{\gamma_j}^{-1}u_{\alpha_j}$.
This is in the form where we can apply Corollary~\ref{property2}, see the
appendix, to evaluate the sums on $z_1$ and $z_2$,
\beq
\begin{split}
\frac{1}{|\Gamma{}|^t}\sum_{u_{b} \in \Gamma{}}
\chi_{R}(d_p) \chi_{R}^*(d_{q'}) = \frac{1}{(2s+1)^2}
\frac{1}{|\Gamma{}|^{t-2}}\sum_{u \in \Gamma{}}
&\chi_{R}(u_{l_4}u_{y_2}u_{l_3}u_{y_1}
u_{l_2}u_{x_2}u_{l_1}u_{x_1}) \\
&\times 
\chi_{R}^* (u_{l_8}u_{y_2}u_{l_7}u_{y_1}u_{l_6}u_{x_2}u_{l_5}u_{x_1}) \ .
\end{split}
\eeq
We can return to the original variables by including two dummy 
variables to account for the sums on $z_1$ and $z_2$. Taking $c_R=\pm 1$ 
for $R(\Gamma), \Gamma \subset \SU(2)$, into account we have thus shown that
\beq\label{eq:double}
\frac{1}{|\Gamma{}|^t}\sum_{u_{b} \in \Gamma{}}
\chi_{R}(d_p) \chi_{R}^*(d_{q'}) = \left(\frac{c_R}{2s+1}\right)^2
\frac{1}{|\Gamma{}|^{t}}\sum_{u \in \Gamma{}}
\chi_{R}(d_p) \chi_{R}^*(d_{q}) \ .
\eeq
%
%
%
Theorem~\ref{maintheorem} now follows by induction, since any orbit $q$ 
can be constructed from $p$ by permuting sections of the orbit 
between a pair of self-intersections or reversing a loop at a 
self-intersection.  \eprf

\begin{figure}[htb]
\begin{center}
\includegraphics[width=3cm]{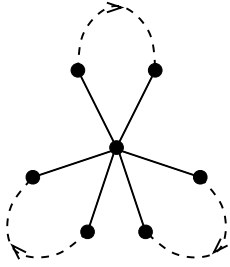}
\end{center}
\caption{An orbit with a degenerate self-intersection.}
\label{fig:loops}
\end{figure}

We comment that at first sight it might not be obvious that all diagrams 
can be constructed via a combination of the two procedures described 
previously. Degenerate cases, in which a system of loops visits the same 
self-intersection more than once, allow both types of reordering.
See Figure~\ref{fig:loops} for an example.

To distinguish the cases it is necessary to follow the orbit 
counting each intersection when it is reached after determining 
whether the order of arcs at the intersection has been changed.  
The number of self-intersections for a given diagram is then $n$. (Note our 
multiple counting of degenerate self-intersections differs from the definition
in \cite{paper:berkolaikoschanzwhitney2}.)

It should also be noted that neither self-retracing loops or repeated bonds
cause difficulties when evaluating the spin contribution.  A self-retracing
section of an orbit $p=(\dots,b_1,b_2,\overline{b_2},\overline{b_1},\dots)$
does not contribute to $d_p$ as $u_{b_1}u_{b_2}u_{b_2}^{-1}u_{b_1}^{-1}=\UI$. 
Similarly, repeated bonds can be removed by a change of variables.

\subsection*{Acknowledgments}
We would like to thank Jonathan Robbins for discussions.
This work has been supported by the European Commission under the 
Research Training Network (Mathematical Aspects of Quantum Chaos) 
no. HPRN-CT-2000-00103 of the IHP Programme.

\section*{Appendix: Group theoretic properties}\label{s:lemmas}

The calculation of spin correlations uses the following properties of 
group representations.
\begin{property}\label{property1a}
Let $R(\Gamma)$ be a unitary irreducible representation of the finite 
group $\Gamma$, and let $a\in \Gamma$.  Then 
\bdm
\frac{1}{|\Gamma|}\sum_{g\in \Gamma} \chi_{R} ( a g^2 )  
=\frac{c_{R}}{\eta{}_R} \chi_{R}(a) \ ,
\edm
where
\bdm
c_R=\left\{\begin{array}{ccl}
1&& R \textrm{ is real} \\
-1&& R \textrm{ is quaternionic}\\
0&& R \textrm{ is not equivalent to $R^*$}\\
\end{array} \right. \ .
\edm
Here $R^*$ is the complex-conjugate representation, $\eta_R$ is the dimension 
of $R(\Gamma)$, and $|\Gamma|$ is the order of the group.
\end{property}
For a proof see Hamermesh \cite{book:hamermesh}, section 5-5. 
This lemma will be applied in an alternate form.
\begin{corollary}\label{property1}
For such a unitary irreducible representation $R(\Gamma)$, 
and $x,y\in \Gamma$,
\bdm
\frac{1}{|\Gamma|}\sum_{g\in \Gamma} \chi_{R} ( x g y g )  
=\frac{c_{R}}{\eta_R} \chi_{R}(xy^{-1}) \ .
\edm
\end{corollary}
\prf In lemma \ref{property1a} replace $g$ with $yg$ and let $x=ay$.  
Summing over $yg$ is equivalent to a sum over $g$ for a fixed element 
$y\in \Gamma$. \eprf

\begin{property}\label{property2a}
Let $R(\Gamma)$ be an irreducible representation of the finite group 
$\Gamma$, and let $x,y\in \Gamma$.  Then 
\bdm
\frac{1}{|\Gamma|} \sum_{g\in \Gamma{}} 
\chi_{R}(xgyg^{-1}) = \frac{1}{\eta_R} \chi_{R}(x) \chi_{R}(y) \ .
\edm
\end{property}

\prf The matrix $\sum_{g\in\Gamma} R(g)R(y)R(g^{-1})$ commutes 
with all elements
of $R(\Gamma)$.  For an irrep $R$, by Schur's lemma,
\beq\label{prop2schur}
\sum_{g\in\Gamma} R(g)R(y)R(g^{-1})=\lambda{} \UI \ .
\eeq
Taking traces yields $|\Gamma{}|\chi_{R}(y)=\eta_R\lambda$.
%
%
Multiplying (\ref{prop2schur}) by $R(x)$ we find,
\beq\label{prop2schurb}
\sum_{g\in\Gamma} R(x)R(g)R(y)R(g^{-1})=\frac{|\Gamma{}|}{\eta_R} R(x) \ .
\eeq
Then taking traces establishes the second lemma.\eprf

\begin{property}\label{property3a}
Let $R(\Gamma)$ be an irreducible representation of the finite group 
$\Gamma$, and let $x,y\in \Gamma$.  Then
\bdm
\frac{1}{|\Gamma|} \sum_{g\in \Gamma{}} \chi_{R}(xg) \chi_{R}(yg^{-1}) 
= \frac{1}{\eta_R} \chi_{R}(xy) \ .
\edm
\end{property}

\prf The matrix $\sum_{g\in\Gamma} R(g)XR(g^{-1})$ commutes 
with all elements
of $R(\Gamma)$ for any matrix $X$.  Again by Schur's lemma,
\beq\label{prop3schur1}
\sum_{g\in\Gamma} R(g)XR(g^{-1})=\lambda{} \UI \ .
\eeq
Take $X$ to have all elements zero except $X_{lm}=1$ and let 
$\lambda=\lambda_{lm}$.
Equation (\ref{prop3schur1}) reads
\beq\label{prop3schur2}
\sum_{g\in\Gamma} R_{il}(g)R_{mj}(g^{-1})=\lambda_{lm} \delta_{ij} \ .
\eeq
Setting $i=j$ and summing over $i$,
\beq
\lambda_{lm}=\frac{|\Gamma{}|}{\eta_R} \delta_{lm}\ .
\eeq
Consequently,
\beq\label{prop3schur3}
\frac{1}{|\Gamma{}|}\sum_{g\in\Gamma} R_{il}(g)R_{mj}(g^{-1})
= \frac{1}{\eta_R} \delta_{lm} \delta_{ij} \ .
\eeq
Multiplying by $R_{li}(x)R_{jm}(y)$
\beq\label{prop3-1}
\frac{1}{|\Gamma{}|}\sum_{g\in\Gamma} R_{li}(x)R_{il}(g)
R_{jm}(y) R_{mj}(g^{-1})
= \frac{1}{\eta_R} R_{li}(x)R_{jm}(y)\delta_{lm} \delta_{ij} \ .
\eeq
Summing on $i$ and $m$, 
\beq\label{prop3-2}
\frac{1}{|\Gamma{}|}\sum_{g\in\Gamma} R_{ll}(xg)R_{jj}(yg^{-1})
= \frac{1}{\eta_R} R_{lj}(x)R_{jl}(y) \ .
\eeq
Finally summing on $j$ and $l$ we obtain lemma \ref{property3a}.\eprf

Combining lemmas \ref{property2a} and \ref{property3a} we obtain a 
corollary that is useful when calculating spin correlations.
\begin{corollary}\label{property2}
For $R(\Gamma)$ an irreducible representation of the finite group 
$\Gamma$, and $a,b,c,d \in \Gamma$,
\bdm
\frac{1}{|\Gamma|} \sum_{g,h\in \Gamma{}} 
\chi_{R}(gah^{-1}bg^{-1}chd) = \frac{1}{\eta_R^2} \chi_{R}(cbad) \ .
\edm
\end{corollary}

{\small
\bibliography{../../../reffs/papers.bib,../../../reffs/books.bib}

\begin{thebibliography}{10}

\bibitem{paper:berkolaiko}
G.~Berkolaiko.
\newblock 
\newblock {\em In this volume}, 2005.

\bibitem{paper:berkolaikoschanzwhitney2}
G.~Berkolaiko, H.~Schanz, and R.~S. Whitney.
\newblock Form factor for a family of quantum graphs: An expansion to third
  order.
\newblock {\em J. Phys. A: Math. Gen.}, 36:8373--8392, 2003.

\bibitem{paper:berkolaikoschanzwhitney}
G.~Berkolaiko, H.~Schanz, and R.~S. Whitney.
\newblock Leading off-diagonal correction to the form factor of large graphs.
\newblock {\em Phys. Rev. Lett.}, 82:104101, 2002.

\bibitem{paper:berry}
M.~V. Berry.
\newblock Semiclassical theory of spectral rigidity.
\newblock {\em Proc. Roy. Soc. Lond. A}, 400:229--251, 1985.

\bibitem{paper:bohigasgiannonischmit}
O.~Bohigas, {M.-J.} Giannoni, and C.~Schmit.
\newblock Characterization of chaotic quantum spectra and universality of level
  fluctuation laws.
\newblock {\em Phys. Rev. Lett.}, 52:1--4, 1984.

\bibitem{paper:bolteharrison}
J.~Bolte and J.~M. Harrison.
\newblock Spectral statistics for the {D}irac operator on graphs.
\newblock {\em J. Phys. A: Math. Gen.}, 36:2747--2769, 2003.

\bibitem{paper:bolteharrison2}
J.~Bolte and J.~M. Harrison.
\newblock The spin contribution to the form factor of quantum graphs.
\newblock {\em J. Phys. A: Math. Gen.}, 36:L433--L440, 2003.

\bibitem{paper:boltekeppeler3}
J.~Bolte and S.~Keppeler.
\newblock Semiclassical form factor for chaotic systems with spin
  $\frac{1}{2}$.
\newblock {\em J. Phys. A: Math. Gen.}, 32:8863--8880, 1999.

\bibitem{book:haake}
F.~Haake.
\newblock {\em Quantum Signatures of Chaos}.
\newblock Springer, 2nd edition, 2001.

\bibitem{book:hamermesh}
M.~Hamermesh.
\newblock {\em Group Theory and its Application to Physical Problems}.
\newblock Addison-Wesley, 1962.

\bibitem{paper:heusler}
S.~Heusler.
\newblock The semiclassical origin of the logarithmic singularity in the
symplectic form factor.
\newblock {\em J. Phys. A: Math. Gen.}, 34:L483-L490, 2001.

\bibitem{paper:keppelermarklofmezzadri}
S.~Keppeler, J.~Marklof, and F.~Mezzadri.
\newblock Quantum cat maps with spin $1/2$.
\newblock {\em Nonlinearity}, 14:719--738, 2001.

\bibitem{paper:kostrykinschrader}
V.~Kostrykin and R.~Schrader.
\newblock Kirchoff's rule for quantum wires.
\newblock {\em J. Phys. A: Math. Gen.}, 32:595--630, 1999.

\bibitem{paper:kottossmilansky2}
T.~Kottos and U.~Smilansky.
\newblock Quantum chaos on graphs.
\newblock {\em Phys. Rev. Lett.}, 79:4794, 1997.

\bibitem{paper:kottossmilansky}
T.~Kottos and U.~Smilansky.
\newblock Periodic orbit theory and spectral statistics for quantum graphs.
\newblock {\em Ann. Phys. (N.Y.)}, 274:76, 1999.

\bibitem{paper:kramers}
H.~A.~Kramers.
\newblock Th\'eorie g\'en\'erale de la rotation paramagn\'etique dans le
cristaux.
\newblock {\em Proc. Acad. Amst.} 33:959--972, 1930.

\bibitem{book:mehta}
M.~L.~Mehta.
\newblock {\em Random Matrices}.
\newblock Academic Press, 1991.

\bibitem{paper:mulleretal}
S.~M\"uller, S.~Heusler, P.~Braun, F.~Haake and A.~Altland.
\newblock Periodic orbit theory of universality in quantum chaos.
\newblock {\em Phys. Rev. E}, 72:046207, 2005.

\bibitem{paper:scharfdietzkushaakeberry}
R.~Scharf, B.~Dietz, M.~Kus, F.~Haake and M.~V.~Berry.
\newblock Kramers' degeneracy and quartic level repulsion.
\newblock {\em Europhys. Lett.}, 5:383--389, 1988.

\bibitem{paper:sieber}
M.~Sieber.
\newblock Leading off-diagonal approximation for the spectral form factor for
  uniformly hyperbolic systems.
\newblock {\em J. Phys. A: Math. Gen.}, 35:L613--L619, 2002.

\bibitem{paper:sieberrichter}
M.~Sieber and K.~Richter.
\newblock Correlations between periodic orbits and their r\^ole in spectral
  statistics.
\newblock {\em Physica Scripta}, T90:128, 2001.

\bibitem{paper:wigner}
E.~P.~Wigner.
\newblock \"Uber die Operation der Zeitumkehr in der Quantenmechanik.
\newblock {\em Nachrichten der Gesellschaft der Wissenschaften zu 
 G\"ottingen. Mathematisch-Physikalische Klasse}, 546--559, 1932. 



\end{thebibliography}
}

\end{document}